\documentclass[prb,aps,showpacs,floatfix,floats,nobalancelastpage,twocolumn]{revtex4}
\usepackage{amsmath, euscript}
\usepackage{amsfonts}
\usepackage{amssymb}
\usepackage{graphicx}
\usepackage{color}
\usepackage{dcolumn}
\usepackage{bm}
\usepackage{subfigure}
\usepackage{ulem}
\def\ra{\rangle}
\def\la{\langle}
\def\up{\uparrow}
\def\dn{\downarrow}
\def\hc{{\rm h.c.}}
\def\b0{{\bf 0}}
\def\bj{{\bf j}}
\def\bJ{{\bf J}}
\def\bk{{\bf k}}

\def\bq{{\bf q}}
\def\br{{\bf r}}
\def\bS{{\bf S}}
\def\bs{{\bf\mbox{\boldmath$\sigma$}}}

\newcommand{\ket}[1]{\left\vert#1\right\rangle}
\newcommand{\eq}[1]{\begin{eqnarray}#1\end{eqnarray}}

\begin{document}

\title{Theory of atomistic simulation of spin-transfer torque in nanomagnets}

%\author{Tiamhock Tay (\begin{CJK}{UTF8}{gbsn}郑添福\end{CJK})}
\author{Tiamhock Tay}
\author{L. J. Sham}

\affiliation{Department of Physics, University of California, San Diego, CA 92093}

\date{\today}


%%%%%%%%%%%%%%%%%%%%%%%%%%%%%%%%%%%%%%%%%%%%%%%%%%%%%%%%%%%%%%%%%%%%%%%%%%%%%%%%%
%%%%%%%%%%%%%%%%%%%%%%%%%%%%%%%%%%%%%%%%%%%%%%%%%%%%%%%%%%%%%%%%%%%%%%%%%%%%%%%%

\begin{abstract}
In spin-transfer torque (STT) for technological applications, the miniaturization of the magnet may reach the stage of requiring a fully quantum-mechanical treatment. We present an STT theory which uses the quantum macrospin ground and excited (magnon) states of the nanomagnet. This allows for energy and angular momentum exchanges between the current electron and the nano-magnet. We develop a method of magnetization dynamics simulation which captures the heating effect on the magnet by the spin-polarized current and the temperature-dependence in STT. We also discuss the magnetostatics effect on magnon scattering for ferromagnetic relaxation in a thin film. Our work demonstrates a realistic step towards simulation of quantum spin-transfer torque physics in nano-scale magnets.
\end{abstract}
\maketitle

%%%%%%%%%%%%%%%%%%%%%%%%%%%%%%%%%%%%%%%%%%%%%%%%%%%%%%%%%%%%%%%%%%%%%%%%%%%%%%%%%
%%%%%%%%%%%%%%%%%%%%%%%%%%%%%%%%%%%%%%%%%%%%%%%%%%%%%%%%%%%%%%%%%%%%%%%%%%%%%%%%%
%%%%%%%%%%%%%%%%%%%%%%%%%%%%%%%%%%%%%%%%%%%%%%%%%%%%%%%%%%%%%%%%%%%%%%%%%%%%%%%%%
%%%%%%%%%%%%%%%%%%%%%%%%%%%%%%%%%%%%%%%%%%%%%%%%%%%%%%%%%%%%%%%%%%%%%%%%%%%%%%%%%
\section{Introduction}

In an early study of the angular momentum transfer between a spin-polarized electron current and a ferromagnetic thin film,\cite{Berger1996} Berger modeled the ferromagnet as a rigid classical spin which effectively acts as a spin-splitting field for the electrons. The torque acting on the magnet is then obtained by appealing to the classical angular momentum conservation law.\cite{Slonczewski1996,bazaliy1998} Berger acknowledged that this semi-classical treatment fails to capture the transverse quantum fluctuations. A recent study by Wang and Sham demonstrated how a fully quantum-mechanical treatment of spin-transfer torque is possible by considering an exchange interaction between each electron and the magnetization,\cite{Wang2012} the latter being modeled by a macrospin. Indeed, their study shows that restoring the transverse correlations between the magnet spins and the current spins gives rise to a noise in the magnetization dynamics of the magnet. This noise may become more important as the recording bits in magnetic storage media approach the nano-scale regime.

The goal of this work is to develop a theory of spin-transfer torque (STT) dynamics and a method of simulation suited to magnets down to nanometer size where the classical treatment of the magnet may fail.\cite{Berger1996,Slonczewski1996,bazaliy1998} We adopt the all-quantum approach of Wang and Sham\cite{Wang2012} in treating both the current electrons and the magnet spins as quantum objects and, in addition, extend the treatment of the magnetization states beyond the rigid macro-spin states to include their excited states as magnons. The interaction between the current electrons and the macrospin ground states involves only elastic scatterings\cite{Berger1996,Slonczewski1996,bazaliy1998} whereas the inclusion of magons accounts for energy transfer which is relevant to the thermal effects.\cite{Katine2000}  Our integration of the magnon dynamics with the macrospin dynamics includes the correlation between the transverse magnet motion and the current electrons, absent in the common magnon treatments in spin torque transfer\cite{Berger1996} and the applications in magnonics.\cite{kruglyak2010} We have also extended the primitive model of a delta function potential for the ferromagnet\cite{Wang2012} to a slab to provide a simple model of dynamics inside the ferromagnet for properties such as the thickness dependence. Thus, our theory provides a basis for an atomistic simulation of the quantum spin-transfer torque process for nano-magnets and the dissipation effects of the magnons on the magnetization dynamics.

An outline of the elements of our theory and key results is as follows. In Sec.~II, the Wang-Sham macrospin approach is applied to a model with a finite thickness for the ferromagnet to furnish the limiting case of our general theory where magnons are neglected. Sec.~III contains an introduction of the magnon to the full Hamiltonian model for spin current driven magnetization dynamics and a description of the application of distorted-wave Born approximation to the current electron scattering with the localized spins of the magnet. The addition of the magnons in the scattering leads to energy as well as angular momentum exchanges between the current and the magnet. Sec.~IV contains the formulation of a theory of wide-angle magnetization precession in an external field. Our theory of the magnetization-magnon interaction resolves the lack of relaxation mechanism in an insulating ferromagnetic film.\cite{Anderson1955,Clogston1956,Arias1999} Using Monte Carlo simulations to study the wide-angle dynamics, we find that magnetic impurity-induced damping results in deviation from the Landau-Lifshitz-Gilbert dynamics. Sec.~V puts the nature of our theory and results in the perspective of the spin-torque field, especially in regard to future work along the concept of this paper.

%%%%%%%%%%%%%%%%%%%%%%%%%%%%%%%%%%%%%%%%%%%%%%%%%%%%%%%%%%%%%%%%%%%%%%%%%%%%%%%%%
%%%%%%%%%%%%%%%%%%%%%%%%%%%%%%%%%%%%%%%%%%%%%%%%%%%%%%%%%%%%%%%%%%%%%%%%%%%%%%%%%
%%%%%%%%%%%%%%%%%%%%%%%%%%%%%%%%%%%%%%%%%%%%%%%%%%%%%%%%%%%%%%%%%%%%%%%%%%%%%%%%%
%%%%%%%%%%%%%%%%%%%%%%%%%%%%%%%%%%%%%%%%%%%%%%%%%%%%%%%%%%%%%%%%%%%%%%%%%%%%%%%%%
\section{Macrospin model without magnons}\label{sec:Yong-model}

We first establish the limiting cases of Refs.~\onlinecite{Berger1996,Slonczewski1996,bazaliy1998,Wang2012} by using Wang and Sham's approach in the coherent scattering of an itinerant electron by a thin ferromagnetic junction in one-dimension. The results of the scattering will be shown to reduce to the classical magnet case when the recoil terms are excluded. This work gives a clearer perspective of Wang and Sham's theory and serves as a limiting case test of our full treatment of spin torque including spin wave excitations due to scattering by the itinerant electrons. Instead of a spin-spin interaction in a $\delta$-plane, we use the following Heisenberg exchange model:
\eq{
  H_{sd} &=& H_{el} + H_{int},\label{eq:full-hamiltonian}
  \\H_{el} &=& \int d^3\br ~c^\dagger_{\br\alpha} \left[-\frac{\hbar^2\nabla^2}{2m} - \lambda_0(\br)\right]c_{\br\alpha},\label{eq:H-el}
  \\H_{int} &=& -\int d^3\br ~\lambda(\br) ~\bj(\br)\cdot\bs_{\alpha\beta} c^\dagger_{\br\alpha} c_{\br\beta},\label{eq:Hint}
  \\&\approx& -\frac{1}{2J} \int d^3\br ~\lambda(\br) ~\bJ\cdot\bs_{\alpha\beta} c^\dagger_{\br\alpha} c_{\br\beta},\label{eq:Hint-macrospin}
}
where summation over each repeated index is implied, $c_{\br\alpha}$ are electron operators with spin component $\alpha$, $\bs_{\alpha\beta}$ are Pauli matrices, and $\lambda(\br)=\lambda$ and $\lambda_0(\br)=\lambda_0$ in the ferromagnet but zero elsewhere. The magnet is described by a continuum background of physically reasonable spin density. Thus, $\bj(\br)$ in Eq.~(\ref{eq:Hint}) represents a coarse-grained spin-1/2 operator in the magnet. By replacing $N\bj(\br)$ with the total spin operator $\bJ$, where $N=2J$ is the total number of localized moments in a fully-saturated magnet, Eq.~(\ref{eq:Hint-macrospin}) gives a macrospin model without spinwaves. In the limit of a macroscopic spin quantum number $J$, $\bJ$ may be replaced by a classical vector and Eq.~(\ref{eq:Hint-macrospin}) would then describe the interaction of an electron spin with an effective magnetic field $\lambda$ in the ferromagnetic film.

The scattering can be block-diagonalized in the total angular momentum basis of the electron spin and the macrospin:
\eq{
  \bS\cdot\bJ &\cong& \left\{
  \begin{array}{ll}
    +J/2, & F=J+1/2
    \\-J/2, & F=J-1/2,
  \end{array}
  \right.
}
where $\bS$ is the electron spin operator and $F$ is the quantum number of the total angular momentum operator ${\bf F}=\bJ+\bS$. Thus, it is more convenient to analyze the scattering in the respective total-spin sector
\eq{
  H_\sigma &=& \int d^3\br ~\psi_\sigma^\dagger(\br) \left\{-\frac{\hbar^2\nabla^2}{2m} - \mu - \frac{\sigma}{2}\lambda(\br)\right\} \psi_\sigma(\br),~~\label{eq:total-spin-Hamiltonian}
}
where $\sigma=\pm$ labels an electron wavefunction in the respective total angular momentum state with $F=J+\sigma/2$. The interacting spin problem now reduces to a single-particle scattering problem. For convenience, we define the Berger frame as one in which the $\hat{z}$-axis (Berger axis) lies along the initial magnetization direction of the magnet prior to scattering by the electron. Consider a given input state in which an electron, with its spin pointing in the direction ($\theta$, $\phi$), moves to the right toward the ferromagnetic film. The scattering problem can be solved by first using Clebsch-Gordon coefficients to transform the spin basis to the total spin basis. After obtaining a scattering solution in the latter basis, we transform back to the former basis.

For the following input state,
\eq{
  \ket{\Psi_{in}} &=& \sum_{\alpha=\pm 1/2} \chi_\alpha \ket{k;\alpha} \otimes \ket{J,J},
}
we obtain an output state to the leading order in $1/\sqrt{J}$ expansion,
\eq{
  \ket{\Psi_{out}} &\approx& \sum_{k'=\pm k} \chi_\dn f_-(k',k)\ket{k';\dn}\otimes\ket{J,J} \nonumber
  \\&+& \sum_{k'=\pm k} \ket{k';\up} \otimes \Bigg\{\chi_\up f_+(k',k)\ket{J,J} \nonumber
  \\&+& \chi_\dn \frac{f_+(k',k) - f_-(k',k)}{\sqrt{2J}}\ket{J,J-1}\Bigg\},\label{eq:Yong}
}
where $f_{\sigma}(k',k)$ is the scattering matrix element ({\it not scattering amplitude}) for an electron in the total spin $F=J+\sigma/2$ to scatter from momentum $k$ to $k'$, cf. Eq.~(\ref{eq:cl-ansatze}). This output state entangles the electron spatial and spin states, and the macrospin states of the magnet. If two successive current electrons are uncorrelated prior to the encounter with the magnet, then the first scattered electron  will decohere into a population distribution of specific electron momentum and spin states which presents an associated mixed state of the magnet for the scattering of the second electron. This introduces noise in the magnetization dynamics of the magnet.\cite{Wang2012}

In the infinite-$J$ limit, the $1/\sqrt{J}$ term drops off and Eq.~(\ref{eq:Yong}) reduces to Berger's solution for an electron scattering off a rigid magnet. This is not unexpected since the Hamiltonian in Eq.~(\ref{eq:total-spin-Hamiltonian}) is formally identical to the semiclassical counterpart if one replaces $\sigma=\pm$ by $\up\dn$. In the classical approximation of the magnetization,\cite{Berger1996,Slonczewski1996} there is no recoil motion in the scattering and the recoil has to be inferred from the conservation of total angular momentum of the current electron and the magnetization. The important consequence of the quantum treatment is that the recoil motion of the magnetization is present in the scattering terms of order $1/\sqrt{J}$. By comparing the second term with a spin coherent state [c.f. Eq.~(\ref{eq:HP2})],\cite{Arecchi1972,Zhang1990} the recoil in the direction of the magnet can be found for a given electron momentum and spin state.

\begin{figure}
  \centering
  \includegraphics[width=3.2in]{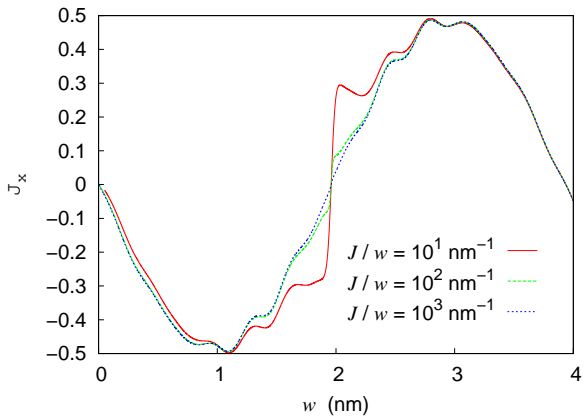}
  \caption{[Color online] For a magnetic film with a given cross section, the $x$-component of the macrospin recoil oscillates as a function of the film thickness when a single electron (initially oriented in the $-\hat{x}$ direction) scatters from the magnet. The recoil shown in the plot has been averaged over all the outcomes in Eq.~(\ref{eq:Yong}). The result for $J/w\geq 10^3 {\rm nm}^{-1}$ is visually indistinguishable from that obtained using Berger's treatment. At $w\approx 2$~nm, quantum-classical differences are noticeable at the mesoscopic range $J\sim 200$ and prominent at the microsopic range of $J \sim 20$.}
  \label{fig:cl-qm-deviation}
\end{figure}

In Fig.~\ref{fig:cl-qm-deviation}, we use Monte Carlo simulations to study the effects of varying $J$ and film thickness on the macrospin recoil. Using a model with a background spin density of $100~{\rm nm}^{-3}$, we compute for different film cross sectional area, the $x$-component of the macrospin recoil as a function of the film thickness after a single electron (initially oriented in the $-\hat{x}$ direction) scatters from the magnet. For simplicity, we replace the discrete-valued $J$ by a continuum $J\geq0.5$ in this calculation. The recoil is averaged over all outcomes in Eq.~(\ref{eq:Yong}). At $w\approx 2$~nm where $f_+\ll f_-$, we observe an appreciable difference between Yong-Sham's macrospin recoil and that of Berger when $J\ll 10^3$. Thus, semiclassical treatments including micromagnetics remain valid for the mean magnetization dynamics down to the middle of the mesoscopic regime\cite{Wernsdorfer2001} but are not able to account for the noise of quantum origin.\cite{Wang2012} The thickness dependence shows that our quantum mechanical electron-magnet scattering is capable of capturing the band structure information in a multilayer device.

%%%%%%%%%%%%%%%%%%%%%%%%%%%%%%%%%%%%%%%%%%%%%%%%%%%%%%%%%%%%%%%%%%%%%%%%%%%%%%%%%
%%%%%%%%%%%%%%%%%%%%%%%%%%%%%%%%%%%%%%%%%%%%%%%%%%%%%%%%%%%%%%%%%%%%%%%%%%%%%%%%%
%%%%%%%%%%%%%%%%%%%%%%%%%%%%%%%%%%%%%%%%%%%%%%%%%%%%%%%%%%%%%%%%%%%%%%%%%%%%%%%%%
%%%%%%%%%%%%%%%%%%%%%%%%%%%%%%%%%%%%%%%%%%%%%%%%%%%%%%%%%%%%%%%%%%%%%%%%%%%%%%%%%
\section{Spin-transfer torque with magnons}\label{sec:HP-model}

For a realistic treatment of the interaction between an itinerant electron and a nanomagnet, we add to the rigid macro-spin states of the magnet excited states approximated by magnons. In this section, we will first state the Holstein-Primakoff (HP) transformation and then relate the correspondence between a rotated ferromagnetic ground state and a bosonic coherent state in the HP representation. We then introduce the Hamiltonian for the electron-ferromagnet system. By analyzing the interaction terms in the HP model, we identify the term corresponding to Berger's scattering potential and terms that provide quantum corrections to the electron-magnet scattering solution. In the distorted-wave Born's approximation, we show that the solution of the preceding section is exactly reproduced by the HP model when spin wave excitations are neglected. The Born approximation is justified by the small change in the excited magnetic state caused by one itinerant electron to the order of $1/\sqrt{J}$.

%%%%%%%%%%%%%%%%%%%%%%%%%%%%%%%%%%%%%%%%%%%%%%%%%%%%%%%%%%%%%%%%%%%%%%%%%%%%%%%%%
%%%%%%%%%%%%%%%%%%%%%%%%%%%%%%%%%%%%%%%%%%%%%%%%%%%%%%%%%%%%%%%%%%%%%%%%%%%%%%%%%
\subsection{Spin coherent state in HP representation}

The Holstein-Primakoff representation for a spin-$j$ operator is given by\cite{HolsteinPrimakoff1940}
\eq{
  j_- = b^\dagger\sqrt{2j - b^\dagger b}, ~~~ j_+ = j^\dagger_-, ~~~ j_z = j - b^\dagger b, \label{eq:HP-def}
}
where the spin state $|j,j\rangle$ serves as the vacuum state $|0\rangle$ for the boson operators $b$ and $b^\dagger$. In the following, the spin and boson operators used to depict the local spins will be labeled with suffices for their positions or lattice momenta. Without loss of generality in representing the dynamics of the magnetization, we may choose its instantaneous axis to make the HP expansion. References~\onlinecite{Arecchi1972,Zhang1990} show that a macrospin state (Dicke state) may be expanded in terms of the coherent states and vice versa. Then, for $J\gg 1$,  a correspondence between a boson coherent state and the macrospin state is established,
\eq{
  && \ket{\eta} = e^{-\vert\eta\vert^2/2} e^{\eta b^\dagger-\eta^*b}\ket{0},\label{eq:HP1}
  \\&\leftrightarrow& \ket{J;\theta,\phi} = e^{-iJ\phi} \frac{e^{\zeta J_-}}{\left(1+\vert\zeta\vert^2\right)^{J}} \ket{J,J},\label{eq:HP2}~~
  \\&& \eta = \zeta\sqrt{2J}, ~~ \zeta = e^{i\phi}\tan(\theta/2).\label{eq:HP3}
}
The large $J$ requirement is easily satisfied even for a small nanomagnet with $10^5$ spin-1/2 moments. For $2J$ such spins, there are as many macrospin states, $|J, J-1, r\rangle$, $r=1,\ldots,2J$. For a lattice of local spins 1/2, Fourier transforms of these states give the spin waves or magnons $b^\dagger_q|0\rangle$,\cite{bloch1930} with wave vectors $q$. The  $|J, J-1, q=0\rangle\equiv b^\dagger_{q=0}|0\rangle$ state is the state which provides the change in the rigid rotation of the macrospin state $|J, J\rangle$ through a small angle $\theta$. This state will be used to determine the direction of recoil for a nanomagnet after an electron scattering.

%%%%%%%%%%%%%%%%%%%%%%%%%%%%%%%%%%%%%%%%%%%%%%%%%%%%%%%%%%%%%%%%%%%%%%%%%%%%%%%%%
%%%%%%%%%%%%%%%%%%%%%%%%%%%%%%%%%%%%%%%%%%%%%%%%%%%%%%%%%%%%%%%%%%%%%%%%%%%%%%%%%
\subsection{The full electron-magnet Hamiltonian}

We now restore magnons to the electron-magnet model by using the full interaction term Eq.~(\ref{eq:Hint}) and also adding magnon energy to the Hamiltonian. Using the HP representation from Eq.~(\ref{eq:HP-def}), we make a series expansion up to quadratic order in the bosonic operators and obtain the following:
\eq{
  H &=& H_0 + V,
  \\H_0 &=& \sum_{\sigma=\up,\dn} \int d^3\br ~c^\dagger_{\br\sigma}\left\{-\frac{\hbar^2\nabla^2}{2m} - \lambda_0(\br) -\frac{\sigma}{2}\lambda(\br)\right\}c_{\br\sigma}\nonumber
  \\&+& \sum_\bq \omega_\bq b^\dagger_\bq b_\bq,\label{eq:H0}
  - \int d^3\br ~\lambda(\br) \left\{c^\dagger_{\br\up} c_{\br\dn} b^\dagger_\br + c^\dagger_{\br\dn} c_{\br\up} b_\br \right\}\nonumber
  \\&+& \sum_{\sigma=\up,\dn} \sigma \int d^3\br ~\lambda(\br) ~c^\dagger_{\br\sigma} c_{\br\sigma} b^\dagger_{\br} b_\br,\label{eq:magnon-magnon}
}
where $\omega_\bq={\cal J}a^2\bq^2/2$ is taken as a phenomenological magnon dispersion for the ferromagnet, with ${\cal J}a^2$ being the spin stiffness and $a$ the spacing between the spin-1/2 moments which we have coarse-grained. For a given magnet with dimensions $L_x$, $L_y$ and $L_z$, the magnon wave vector has components taking discrete values $q_i=2\pi n_i/L_i$ where $n_i=0,1,\ldots,L_i/a$. We have neglected the magnon-magnon interaction since the electron scattering time is much shorter than the relaxation time of a magnon mode.

\begin{figure}
  \centering
  \begin{tabular}{ccc}
    \subfigure[]{
      \includegraphics[width=1in]{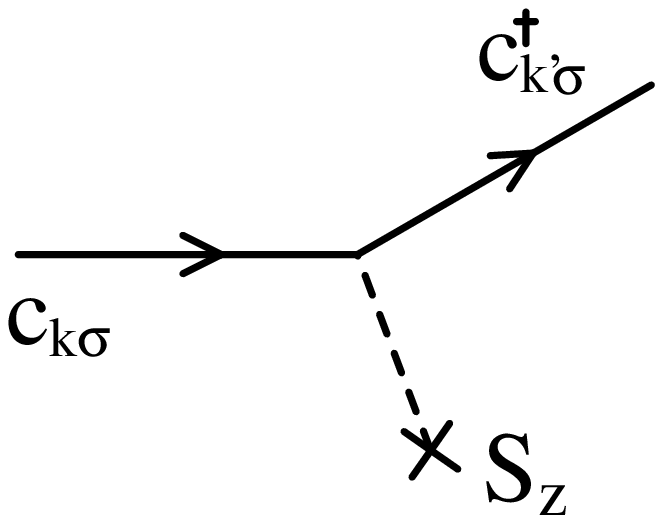}
      \label{fig:feynman-a}
    } &
    \subfigure[]{
      \includegraphics[width=1in]{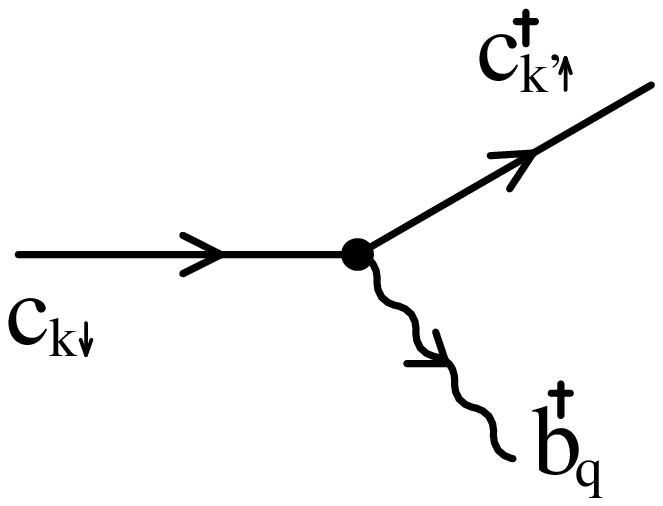}
      \label{fig:feynman-b}
    } &
    \subfigure[]{
      \includegraphics[width=1in]{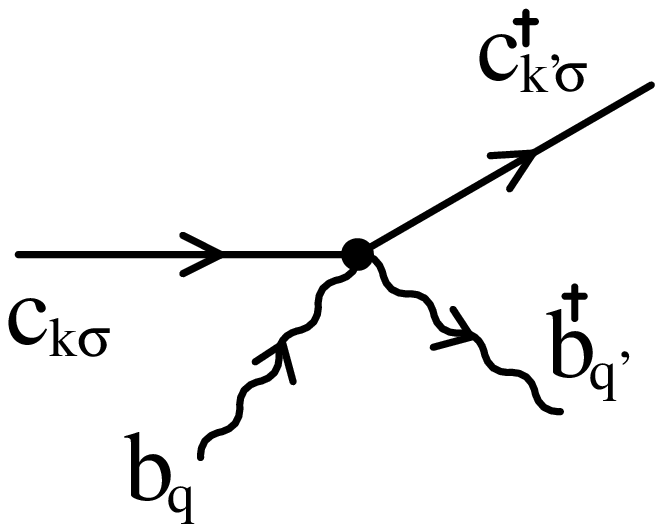}
      \label{fig:feynman-c}
    }
  \end{tabular}
  \caption{Illustration of the interactions between an electron and a nanomagnet in the HP representation. (a) Classical scattering potential in Berger's treatment. (b) Electron spin flip causes orientation recoil of the magnet or spin wave excitations. (c) Electron-magnon interaction results in an effective renormalization of the classical scattering potential, leading to a temperature dependence in STT.}
\end{figure}

The spin-dependent term in $H_0$ comes from the electron-magnet interaction; it gives the classical scattering potential in Berger's theory and may be represented using the Feynman diagram in Fig.~\ref{fig:feynman-a}. The effect of this term is discussed in Sec.~\ref{sec:Yong-model}, where the recoiless scattered state is given by the leading-order term in Eq.~(\ref{eq:Yong}).

The two terms in $V$ are depicted by Figs.~\ref{fig:feynman-b} and \ref{fig:feynman-c}. The cubic term contributes to Berger's scattering solution a quantum correction of the form $\ket{\pm\bk';\up} \otimes b^\dagger_\bq\ket{M}$, where $\ket{M}$ denotes the initial state of the magnet prior to scattering. For $\bq=\b0$, $b^\dagger_\bq\ket{0}\equiv\ket{J,J-1}$ results in a recoil in the orientation of the magnet, and later we will show that this reproduces Wang and Sham's macrospin result exactly in the leading $1/\sqrt{J}$ approximation. For $\bq\neq\b0$, a magnon is created. This leads to a heating of the ferromagnetic state by the spin-polarized  current. The stochastic selection by a subsequent electron of the magnetization states from the elastic and inelastic scattering by the preceding electron  determines the new direction and the internal state of the magnetization.

The second electron-magnon term in $V$ further contributes a quantum correction of the following form
$\ket{\pm\bk';\sigma} \otimes b^\dagger_{\bq'} b_\bq\ket{M}$. For $\bq'=\bq$, the magnon population has the effect of renormalizing the $S_z$-$j_z$ spin coupling between the electron and the magnet and thus contributes to temperature dependence in the orientation recoil. In the rest of the section, we develop a scattering theory to study the effects of the electron-magnet interaction.

%%%%%%%%%%%%%%%%%%%%%%%%%%%%%%%%%%%%%%%%%%%%%%%%%%%%%%%%%%%%%%%%%%%%%%%%%%%%%%%%%
%%%%%%%%%%%%%%%%%%%%%%%%%%%%%%%%%%%%%%%%%%%%%%%%%%%%%%%%%%%%%%%%%%%%%%%%%%%%%%%%%
%\subsection{Keeping $H_0$ only}
\subsection{Classical scattering potential}

This subsection considers the one-dimensional problem of an electron scattering off a spin-dependent classical potential (i.e. without the interaction term $V$) which extends from $x=-w/2$ to $x=w/2$. Scattering solutions that will be used for calculations in later sections are those prepared with (i) an incident electron approaching the magnet from the left, (ii) a reflected electron moving left from the magnet, and (iii) a transmitted electron moving right from the magnet. It is a simple exercise to obtain the following solutions:
\begin{widetext}
  \eq{
    % First block
    &\begin{aligned}[c]
       \Psi^{(+)}_{+k\sigma}(x) &=&
       \left\{
       \begin{array}{lcl}
         e^{ikx} + f_\sigma(-k,k)e^{-ikx} & , ~~~~~~~ & ~-\infty<x<-w/2 \\
         g_\sigma e^{ik_\sigma x} + h_\sigma e^{-ik_\sigma x} & , ~~~~~~~ & -w/2<x<+w/2 \\
         f_\sigma(+k,k) e^{ikx} & , ~~~~~~~ & +w/2<x<+\infty
       \end{array}
       \right.\label{eq:cl-ansatze}\\
       \Psi^{(-)}_{-k\sigma}(x) &=&
       \left\{
       \begin{array}{lcl}
         e^{-ikx} + f^*_\sigma(-k,k)e^{ikx} & , ~~~~~~~ & ~-\infty<x<-w/2 \\
         g^*_\sigma e^{-ik_\sigma x} + h^*_\sigma e^{ik_\sigma x} & , ~~~~~~~ & -w/2<x<+w/2 \\
         f^*_\sigma(+k,k) e^{-ikx} & , ~~~~~~~ & +w/2<x<+\infty
       \end{array}
       \right.\\
       \Psi^{(-)}_{+k\sigma}(x) &=&
       \left\{
       \begin{array}{lcl}
         f^*_\sigma(+k,k) e^{ikx} & , ~~~~~~~ & ~-\infty<x<-w/2 \\
         g^*_\sigma e^{ik_\sigma x} + h^*_\sigma e^{-ik_\sigma x} & , ~~~~~~~ & -w/2<x<+w/2 \\
         e^{ikx} + f^*_\sigma(-k,k) e^{-ikx} & , ~~~~~~~ & +w/2<x<+\infty
       \end{array}
       \right.
     \end{aligned}&\\
    \nonumber\\
    %Second block
    &\begin{aligned}[c]
       f_\sigma(+k,k) &= -4kk_\sigma/c_\sigma,\\
       f_\sigma(-k,k) &= 2i(k^2-k^2_\sigma)\sin(k_\sigma w)/c_\sigma,\\
       \hbar^2 k^2_\sigma/2m &= \hbar^2 k^2/2m + \delta\mu + \sigma\lambda/2,\\
     \end{aligned} ~~~~~
    %Third block
    \begin{aligned}[c]
      g_\sigma &= -2k(k+k_\sigma) e^{i(k-k_\sigma)w/2}/c_\sigma,\\
      h_\sigma &= +2k(k-k_\sigma) e^{i(k+k_\sigma)w/2}/c_\sigma,\\
      c_\sigma &= (k-k_\sigma)^2 e^{i(k+k_\sigma)w} - (k+k_\sigma)^2 e^{i(k-k_\sigma)w}.
    \end{aligned}& \label{eq:cl-ampl}
  }
\end{widetext}
For the state $\Psi^{(+)}_{+k\sigma}(x)$ in Eq.~(\ref{eq:cl-ansatze}), $f_\sigma(\pm k,k)$ give the matrix element for a spin-$\sigma$ electron to be transmitted or reflected from the magnet. $\Psi^{(-)}_{-k\sigma}(x)$ is the complex conjugate of $\Psi^{(+)}_{k\sigma}(x)$, while $\Psi^{(-)}_{k\sigma}(x)$ is the mirror reflection of $\Psi^{(-)}_{-k\sigma}(x)$. In the following sections, these states which are commonly known as the  ``distorted waves'', will serve as zeroth-order wave functions for the first Born approximation of the finite $q$ magnon contributions.

%%%%%%%%%%%%%%%%%%%%%%%%%%%%%%%%%%%%%%%%%%%%%%%%%%%%%%%%%%%%%%%%%%%%%%%%%%%%%%%%%
%%%%%%%%%%%%%%%%%%%%%%%%%%%%%%%%%%%%%%%%%%%%%%%%%%%%%%%%%%%%%%%%%%%%%%%%%%%%%%%%%
\subsection{Energy and angular momentum exchanges}

%%%%%%%%%%%%%%%%%%%%%%%%%%%%%%%%%%%%%%%%%%%%%%%%%%%%%%%%%%%%%%%%%%%%%%%%%%%%%%%%%
\subsubsection{Distorted-wave Born approximation}
Although the exchange coupling constant $\lambda$ which appears in $H_0$ as well as $V$ is not necessarily small, we explain below that it is still reasonable to treat $V$ as a perturbation on $H_0$. 
Averaging over all $N$ localized moments, $\la b^\dagger_\br b_\br\ra\sim n_{tot}/N$ shows that $c^\dagger_{\br\sigma} c_{\br\sigma} b^\dagger_{\br} b_\br$ indeed gives a small contribution when the temperature is much lower than the Curie temperature ($T_C$), since the total magnon population $n_{tot}\ll N$. For $c^\dagger_{\br\dn} c_{\br\up} b_\br$, only a small number of magnon modes with non-zero occupation contribute to the scattering when $T\ll T_C$. The contribution from $c^\dagger_{\br\up} c_{\br\dn} b^\dagger_\br$ is comparable to $\lambda$, but for $T\ll T_F$ (the Fermi temperature), a large number of scattering outcomes with magnon creation are forbidden by Pauli exclusion principle since the outgoing electron has to scatter into a state deep below the Fermi level. %of the normal metal.

Hence, we may treat $V$ as a small perturbation. Approximate scattering solutions can be obtained from the following $T$-matrix formula:\cite{RogerNewton}
\eq{
  T_{\beta\alpha} &\approx& T^{(0)}_{\beta\alpha} + \big\la \Psi^{(-)}_\beta\big\vert V\big\vert\Psi^{(+)}_\alpha\big\ra,~~\label{eq:additive-interaction}
}
where $\alpha$ and $\beta$ label the eigenstates of $H_0$, and $T^{(0)}_{\beta\alpha}$ and $T_{\beta\alpha}$ are $T$-matrix elements for an incoming state $\alpha$ to scatter into outgoing state $\beta$ for respective Hamiltonians $H_0$ and $H_0+V$. Here, $\Psi^{(\pm)}_\alpha$ are exact scattering states of the Hamiltonian $H_0$, with $+(-)$ indicating incoming (outgoing) states. Note that the total energy of the system has to be conserved by including energy exchange between the electron and the magnon states. The first Born's approximation is used to derive the above formula.\cite{RogerNewton}

%%%%%%%%%%%%%%%%%%%%%%%%%%%%%%%%%%%%%%%%%%%%%%%%%%%%%%%%%%%%%%%%%%%%%%%%%%%%%%%%%
\subsubsection{Recoil in orientation of the magnet}

We now determine the change in orientation of the nanomagnet, the magnon creation or annihilation amplitudes for inelastic events after electron decoherence occurs. For the electron-nanomagnet scattering model, we make use of exact expressions for $T^{(0)}_{\beta\alpha}$ and $\left\vert\Psi^\pm_\alpha\right\ra$ from Eq.~(\ref{eq:cl-ansatze}). Using the T-matrix formula, we derive explicit expressions as the starting point to investigate the effect of magnons on the rigid rotation of the magnetization. Here, we retain only the $b_{\bq=\b0}$ term in $V$ while leaving the $\bq\neq 0$ terms to the next section. The Hamiltonian relevant to the scattering between an itinerant electron and the macrospin is
\eq{
  H &=& H_0 - \frac{\lambda}{\sqrt{N}} \int_{-w/2}^{w/2} dx ~ \left(b^\dagger_\b0 c^\dagger_{\br\up} c_{\br\dn} + \hc\right),
}
where the number of spins $N=2J$. Taking overlap for the spatial part using the scattering solutions in Eq.~(\ref{eq:cl-ansatze}), we obtain the following matrix elements:
\eq{
  \la \Psi^{(-)}_{\pm k\up}\vert V\vert\Psi^{(+)}_{k\dn}\ra &=& -\frac{\lambda}{\sqrt{N}} b^\dagger_\b0 \int_{-w/2}^{w/2} dx ~\Psi^{(-)*}_{\pm k\up}(x) \Psi^{(+)}_{+k\dn}(x),\nonumber\\
  \\
  &\equiv& -\frac{\hbar^2k}{im} \frac{f_\up(\pm k,k)-f_\dn(\pm k,k)}{\sqrt{N}} ~b^\dagger_\b0.~~~
}
For an incident electron with spin state $(\chi_\up,\chi_\dn)$, the scattered state of the electron-magnet system is
\eq{
  \ket{\Psi_{out}} &\approx& \sum_{k'=\pm k} \chi_\dn f_\dn(k',k)\ket{k';\dn}\otimes\ket{M} \nonumber\\
  &+& \sum_{k'=\pm k} \ket{k';\up} \otimes \Bigg\{\chi_\up f_\up(k',k)\ket{M} \nonumber\\
  &+& \chi_\dn\frac{f_\up(k',k) - f_\dn(k',k)}{\sqrt{2J}} b^\dagger_\b0\ket{M}\Bigg\}.\label{eq:HP-macrospin}
}
If the magnet is initially in the ferromagnetic ground state where $\ket{M}=\ket{J,J}$ and $b^\dagger_\b0\ket{M}=\ket{J,J-1}$, the HP scattering solution reproduces Eq.~(\ref{eq:Yong}) {\it exactly}. Thus, the distorted-wave Born approximation indeed allows an accurate treatment of the macrospin recoil in the leading $1/\sqrt{N}$ order.

%%%%%%%%%%%%%%%%%%%%%%%%%%%%%%%%%%%%%%%%%%%%%%%%%%%%%%%%%%%%%%%%%%%%%%%%%%%%%%%%%
\subsubsection{Effect of magnons on orientation recoil}

\begin{figure}
  \centering
  \includegraphics[width=\columnwidth]{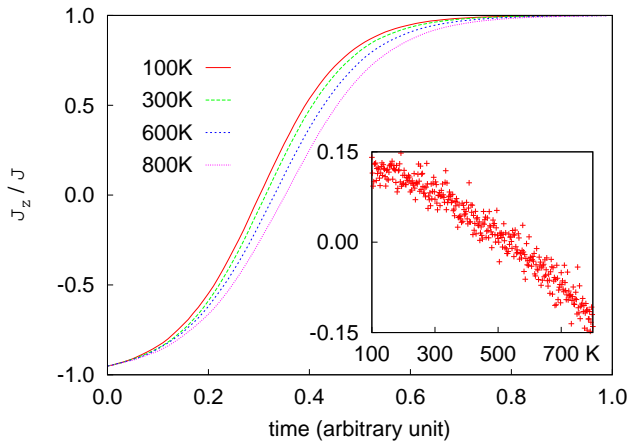}
  \caption{At high temperatures, the reduced total spin of the magnet should lead to a larger orientation recoil as electrons scatter off the magnet. But a smaller difference in the scattering amplitudes for spin up and spin down electrons at high magnon population over-compensates and results in a reduced STT. The inset shows a vertical cut at $t=0.32$, plotted against the magnet temperature. The scatter results from the stochastic nature of individual Monte Carlo runs.}
  \label{fig:HAMR}
\end{figure}

Consider first the two-magnon interaction with the itinerant electron in $V$,
\eq{
    \frac{1}{N} \sum_{\bq_1\bq_2} b^\dagger_{\bq_1} b_{\bq_1+\bq_2} \sum_\sigma \sigma \int d^d\br ~\lambda(\br) e^{i\bq_2\cdot\br} ~c^\dagger_{\br\sigma} c_{\br\sigma}.
}
For $\bq_2=\b0$, the contribution renormalizes the $S_z$-$j_z$ coupling between an itinerant electron and the magnet. Neglecting terms with $\bq_2\neq\b0$, we move the diagonal terms to the non-interacting part of the Hamiltonian $H_0$ and replace the magnon occupations by their mean-field expectation values. This approximation leads to a renormalized, anisotropic spin coupling between the electron and the macrospin:
\eq{
  c^\dagger_{\br\alpha} (\bs_{\alpha\beta} \cdot \bj_\br) c_{\br\beta} &\approx& c^\dagger_{\br\up} c_{\br\dn} j^-_\br \nonumber + c^\dagger_{\br\dn} c_{\br\up} j^+_\br \nonumber\\
  &+& \frac{\sigma}{2} c^\dagger_{\br\sigma} c_{\br\sigma} \left(1 - \sum_\bq \la n_\bq\ra/J\right).~~~\label{eq:ani}
}
The scattered state in Eq.~(\ref{eq:HP-macrospin}) is accordingly extended to
\eq{
  \ket{\Psi_{out}} &\approx& \sum_{\pm} \chi_\dn \tilde{f}_\dn(\pm k,k)\ket{\pm k;\dn}\otimes\ket{M} \nonumber\\
  &+& \sum_{\pm} \ket{\pm k;\up} \otimes \Bigg\{\chi_\up \tilde{f}_\up(\pm k,k)\ket{M} \nonumber\\
  &+& \chi_\dn\frac{\tilde{f}_\up(\pm k,k) - \tilde{f}_\dn(\pm k,k)}{\sqrt{2J}[1-\sum_\bq\la n_\bq\ra/J]} b^\dagger_\b0\ket{M}\Bigg\},\label{eq:HAMR}
}
where $\tilde{f}_\sigma(\pm k,k)$ are the matrix elements of an electron scattering from a magnet, computed using the reduced spin splitting $\lambda[1-\sum_\bq\la n_\bq\ra/J]$. Two factors with competing effects are present in the spin flip term. The extra factor $[1-\sum_\bq\la n_\bq\ra/J]$ in the denominator acts to increase the recoil magnitude at large magnon populations, while the reduced $S_z$-$j_z$ coupling leads to a smaller difference between the spin up and spin down scattering amplitudes. To determine the resulting temperature dependence in magnetization switching, we simulate the trajectories of the magnetization at various temperatures using Monte Carlo. Figure~\ref{fig:HAMR} shows that the reversal rate decreases with increasing temperature, thus allowing us to conclude that STT is weaker in the presence of magnons. Note that this does not conflict with heat-assisted magnetic recording (HAMR) since the latter occurs via a temporary loss of magnetism by heating a magnet beyond its Curie point, followed by remagnetization in a magnetic field as it cools. Note also that we have excluded magnetic and anisotropy fields in this simulation so that the activation effect of an energy barrier on the switching is removed and hence, allowing us to isolate the effect of magnon population on the STT. Therefore, Fig.~\ref{fig:HAMR} should not be confused with ``thermally-activated switching''; the latter pertains to the Arrhenius factor in magnetization reversals.\cite{Brown1963,Koch2000,You2011} Full treatment of STT-driven switching in a field will be deferred to Sec.~\ref{sec:switching-in-field}.

We now discuss the neglect of spinwave transformation under rotation in our formulation. In Eq.~(\ref{eq:HAMR}), one observes that the operator $b^\dagger_\b0$ rotates the ground state of the magnet but not the spinwaves. For small-angle recoils typical of nanomagnets with at least $10^5$ localized spin-1/2 moments, an $O(N^{-1/2})$ correction to the spinwave operators under rotation is indeed negligible. Therefore, we may regard $b^\dagger_\b0\ket{M}$ in Eq.~(\ref{eq:HAMR}) as a rigid rotation of all the spins in the magnet. Alternatively, one may approximate the state of a magnet at finite temperature by $\ket{J',J'}$ with a reduced total spin $J'=J[1-\sum_\bq\la n_\bq\ra/J]$ and analyze the orientation recoil by appealing to the original interaction Hamiltonian:
\eq{
  H_{int} &=& -\sum_{\alpha\beta} \int d^d\br ~\lambda(\br) ~c^\dagger_{\br\alpha} \bs_{\alpha\beta} c_{\br\beta} \cdot \bj_\br,\\
  &\approx& -\frac{\bJ}{N} \cdot \sum_{\alpha\beta} \int d^d\br ~\lambda(\br) ~c^\dagger_{\br\alpha} \bs_{\alpha\beta} c_{\br\beta}\nonumber\\
  &-& \sum_{\bq\neq\b0} \frac{b^\dagger_\bq}{\sqrt{N}} \cdot \int d^d\br ~\lambda(\br) e^{-i\bq\cdot\br} ~c^\dagger_{\br\up} c_{\br\dn} + \hc\nonumber\\
  &+&\sum_{\bk;\bq\neq\b0} \frac{b^\dagger_\bk b_{\bk+\bq}}{N} \cdot \sum_\sigma \sigma \int d^d\br ~\lambda(\br) e^{i\bq\cdot\br} ~c^\dagger_{\br\sigma} c_{\br\sigma}.\nonumber\label{eq:HAMR-magnon}\\
}
The first term in Eq.~(\ref{eq:HAMR-magnon}) causes a rigid rotation of the spinwaves along with the uniform magnetization as an electron scatters off elastically from the magnet. The remaining terms involve inelastic processes and will be discussed next. In Born's approximation, the orientation recoil is analytically identical to the preceding formulation at the leading $1/N$ order. Thus, the HP representation allows one to study magnetization recoils at zero temperature as well as finite temperatures.

%%%%%%%%%%%%%%%%%%%%%%%%%%%%%%%%%%%%%%%%%%%%%%%%%%%%%%%%%%%%%%%%%%%%%%%%%%%%%%%%%
\subsubsection{Magnon creation or annihilation}

The remaining single-magnon and electron interaction terms for the derivation of the coefficients of amplitudes of states with a magnon created or annihilated,\cite{Bonca1995,Emberly2000} are
\eq{
  -\frac{1}{\sqrt{N}} \sum_{\bq\neq\b0} b^\dagger_\bq \int d^d\br ~\lambda(\br) ~c^\dagger_{\br\up} c_{\br\dn} e^{-i\bq\cdot\br} + \hc.
}
As before, we use the scattering solutions in Eq.~(\ref{eq:cl-ansatze}) to obtain the following matrix elements
\eq{
  \la\Psi^{(-)}_{\bk'\up}\vert V\vert\Psi^{(+)}_{\bk\dn}\ra &=& -\frac{\lambda}{\sqrt{N}} \sideset{}{^\prime}\sum_\bq \upsilon_{\up\dn}(\bk',\bk;\bq) ~b^\dagger_\bq,\\
  \la\Psi^{(-)}_{\bk'\dn}\vert V\vert\Psi^{(+)}_{\bk\up}\ra &=& -\frac{\lambda}{\sqrt{N}} \sideset{}{^\prime}\sum_\bq \upsilon_{\dn\up}(\bk',\bk;\bq) ~b_\bq,\\
  \upsilon_{\bar{\sigma}\sigma}(\bk',\bk;\bq) &=& \int_{-w/2}^{w/2} dx ~\Psi^{(-)*}_{\bk'\bar{\sigma}}(x) \Psi^{(+)}_{\bk\sigma}(x) e^{i\sigma q_xx},~~~~
}
where the primed sum indicates restriction of the scatterings to an equal total energy shell. Then the magnon correction to the scattering solution is
\eq{
  \frac{im\lambda}{\sqrt{N}\hbar^2} \sum_{\bq,\bk'} \bigg\{\frac{\chi_\dn}{k'_x(E,\bq)} \upsilon_{\up\dn}(\bk',\bk;\bq)\ket{\bk';\up} \otimes b^\dagger_\bq\ket{M} \nonumber\\
  + ~~\frac{\chi_\up}{k'_x(E,\bq)} \upsilon_{\dn\up}(\bk',\bk;\bq)\ket{\bk';\dn} \otimes b_\bq \ket{M} \bigg\},\label{eq:magnon-state}\\
  k'(E,\bq) = \sqrt{k^2 \mp 2m\omega_\bq/\hbar^2}. \hspace{3.1cm}\label{eq:energy-balance}
}
Equation~(\ref{eq:magnon-state}) shows that a spin-down electron flips upward by emitting a magnon while a spin-up electron flips downward by absorbing a magnon, with the energy balance provided by the kinetic energy of the electron in Eq.~(\ref{eq:energy-balance}).

\begin{figure}
  \centering
  \includegraphics[width=\columnwidth]{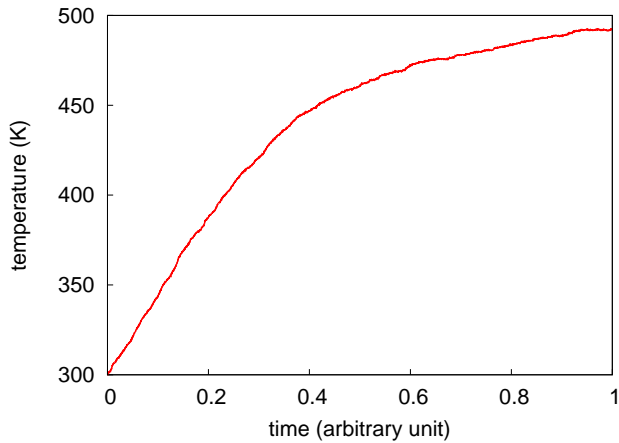}
  \caption{Monte Carlo simulation of the spin lattice temperature during magnetization switching. Due to a large number of magnon modes with zero occupation number, the probability for magnon creation is greater than that for absorption, thus leading to an overall monotonic temperature rise during the switching process.}
  \label{fig:magnon}
\end{figure}

Figure~\ref{fig:magnon} is a simulation of the time evolution of the magnet temperature as a nanomagnet undergoes STT-driven switching. After each inelastic scattering event that occurred during our Monte Carlo simulation, we fit the non-equilibrium distribution of the magnet states by a single effective temperature before the next electron is injected into the magnet. More rigorous treatment of the magnetization dynamics will be left for future studies. The result shows that the spin temperature of the nanomagnet rises monotonically on average, but with a decreasing rate of rise. This is consistent with the scattering result in Eq.~(\ref{eq:magnon-state}). For magnon absorption, the contribution comes only from low energy states with non-zero occupation. On the other hand, all magnon states contribute to magnon creation regardless of their occupation numbers. Thus, the total probability for an electron to emit a magnon is evidently much greater than that for absorbing a magnon. Inclusion of the effect of Pauli exclusion on an electron scattering into states below the Fermi level of the normal metal leads to comparable probabilities for magnon creation and destruction, but generally higher for magnon creation. For relative orientation of the magnetization and electron spin close to the anti-parallel case, $\vert\chi_\dn\vert \gg \vert\chi_\up\vert$ results in more magnon creation than magnon destruction. When the electron spin and the magnetization are nearly parallel, the opposite occurs. However, the destruction of many low-energy magnons together with the creation of fewer high-energy magnons do not necessarily result in a net heat extraction from the magnon bath. This explains the monotonous temperature rise of the magnon bath. On average, less than $5\%$ of the scatterings in our simulation are inelastic; this justifies the distorted-wave Born approximation used. For this simulation, we exclude couplings between the magnon, electronic and phonon degrees of freedom.\cite{Kirilyuk2010} Inclusion of these effects phenomenologically should result in a temperature peak followed by a cooling of the spin lattice during the magnetization reversal.

%%%%%%%%%%%%%%%%%%%%%%%%%%%%%%%%%%%%%%%%%%%%%%%%%%%%%%%%%%%%%%%%%%%%%%%%%%%%%%%%%
%%%%%%%%%%%%%%%%%%%%%%%%%%%%%%%%%%%%%%%%%%%%%%%%%%%%%%%%%%%%%%%%%%%%%%%%%%%%%%%%%
%%%%%%%%%%%%%%%%%%%%%%%%%%%%%%%%%%%%%%%%%%%%%%%%%%%%%%%%%%%%%%%%%%%%%%%%%%%%%%%%%
%%%%%%%%%%%%%%%%%%%%%%%%%%%%%%%%%%%%%%%%%%%%%%%%%%%%%%%%%%%%%%%%%%%%%%%%%%%%%%%%%
\section{Field effects on magnetization dynamics}\label{sec:switching-in-field}

In the preceding sections, we developed a consistent theory of  energy and angular momentum exchanges between an itinerant electron and a magnet by treating the magnet states quantum mechanically. We now turn to the problem of field-driven precession in the intervening time interval between two successive electron scatterings. Consider a fixed reference frame that coincides with the Berger frame at time $t=0$. At any time, the ground state of the magnet is always given by the vacuum state in the instantaneous Berger frame, but relative to the fixed reference frame, $\ket{0}$ transforms into a coherent state $\ket{\eta_\b0(t)}$ defined in Eq.~(\ref{eq:HP1}). By $\eta_\b0(t)=\la\eta_\b0(t)\vert b_\b0\vert\eta_\b0(t)\ra$ and the Heisenberg equation for $b_\b0$, we obtain $\dot{\eta}_\b0(t)$. Although the HP approach is restricted to small-angle dynamics about the quantization axis, the full trajectory of the magnetization can be built up using infinitesimal time evolution in a series of Berger frames. For the purpose of modeling, one might approximate the quantum state of the magnet using a direct product of magnon coherent states. This avoids the need for a prohibitively large amount of resources in simulating the distribution of magnet states. We leave the simulation of the $\bq\neq 0$ magnon evolution for future studies.

%%%%%%%%%%%%%%%%%%%%%%%%%%%%%%%%%%%%%%%%%%%%%%%%%%%%%%%%%%%%%%%%%%%%%%%%%%%%%%%%%
%%%%%%%%%%%%%%%%%%%%%%%%%%%%%%%%%%%%%%%%%%%%%%%%%%%%%%%%%%%%%%%%%%%%%%%%%%%%%%%%%
\subsection{HP approach for wide-angle precession}\label{sec:field}

In addition to the Zeeman energy of a magnet in an external magnetic field, we include the effect of dipolar interaction, crystalline and shape anisotropies by defining an effective field $\vec{h}$ that exerts the same net torque on the magnet.\cite{kittel1948} To treat relaxation in the magnetization dynamics, we include a $b^\dagger_{\b0}b_\bq$ scattering term in our Hamiltonian and obtain the spin relaxation in terms of rates. The starting magnon Hamiltonian is
\eq{
  H_M &=& \frac{1}{\sqrt{N}} \sum_{\bq\neq\b0} W_\bq b^\dagger_\bq b_\b0 + \frac{h_+}{4\sqrt{N}} \sum_{\bq\bq'}b^\dagger_\bq b^\dagger_{\bq'}b_{\bq+\bq'} + \hc\nonumber
  \\&+& \sum_\bq(h_z+\omega_\bq)b^\dagger_\bq b_\bq - \frac{\sqrt{N}}{2}(h_+b^\dagger_\b0+h_-b_\b0),\label{eq:Hamiltonian-in-field}
}
where $h_\pm=h_x\pm ih_y$, $\omega_\bq$ contains contributions from Heisenberg exchange and demagnetization field,\cite{HolsteinPrimakoff1940,Clogston1956} and $W_\bq$ gives the interaction between the uniform mode and the magnons. In Ref.~\onlinecite{Sparks1961}, Sparks {\it et al} showed that the $W_\bq$ term arises from impurity fields and is responsible for damping. The terms containing $h_\pm$ or $h_z$ are obtained by expanding the HP representation of the effective Zeeman energy $-\vec{h}\cdot\bJ$ to cubic order and then transforming to momentum space. We exclude the current electron-magnet interaction from $H_M$ and apply this Hamiltonian in between two successive electron scatterings. In the following, $\eta_\bq$ represents the magnon expectation in the coherent state $\ket{\eta_\bq(t)}$. Taking expectation of the Heisenberg equation of motion for $b_\bq$ with respect to $\ket{\eta_\bq(t)}$, we obtain
\eq{
  i\partial_t \eta_\bq &=& \frac{1}{\sqrt{N}}(1-\delta_{\bq 0})W_\bq\eta_\b0 + \frac{\delta_{\bq 0}}{\sqrt{N}} \sum_{\bq'\neq 0} W^*_{\bq'}\eta_{\bq'}
  \\&+& (h_z+\omega_\bq)\eta_\bq - \frac{\sqrt{N}}{2}h_+\delta_{\bq 0}\nonumber
  \\&+& \frac{h_+}{2\sqrt{N}}\sum_{\bq'}\eta^*_{\bq'}\eta_{\bq+\bq'} + \frac{h_-}{4\sqrt{N}}\sum_{\bq'}\eta_{\bq'}\eta_{\bq-\bq'}.\label{eq:differentialEq}~~~
}
Note that the constant $O(\sqrt{N})$ term results in a strongly driven harmonic oscillator coupled weakly to a reservoir of interacting oscillators. Ignoring all terms except this, we obtain $\partial_t \eta_\b0(t)=i\sqrt{N}h_+/2$. Equation~(\ref{eq:HP1}) gives $\dot{\theta} e^{i\phi}=ih_+$ which results in a trajectory pointing along $\hat{z}\times\vec{h}$ in the Berger frame. In a fixed reference frame, this reproduces the classical result $\partial_t\vec{M}=\vec{M}\times\vec{h}$. 

To obtain the damping effect due to the magnons, we use $\eta_\b0 = \tilde{\eta}_\b0 e^{-i(h_z+\tilde{\omega}_\b0) t}+\sqrt{N}\xi_\b0$ to 
obtain
\eq{
  \dot{\theta} e^{i\phi} &=& i(2h_z+2\tilde{\omega}_\b0-i\Gamma)\xi_\b0,\label{eq:damping}
  \\\Gamma &=& \frac{2\pi}{N} \sideset{}{^\prime}\sum_\bq \big\vert W_{\bq}\big\vert^2 \delta(\tilde{\omega}_\b0-\omega_\bq),\label{eq:energy-conservation}
  \\\tilde{\omega}_\b0 &=& \omega_\b0 + \frac{1 - {\frac{3}{4}}\vert\xi_\b0\vert^2}{(1-\vert\xi_\b0\vert^2)^{3/2}}\vert\xi_\b0 h_+\vert,\label{eq:freq-shift}
}
with $\xi_\b0=\vert\xi_\b0/h_+\vert h_+$ satisfying
\eq{
  \frac{4\vert\xi_\b0\vert^2-3\vert\xi_\b0\vert^4}{2(1-\vert\xi_\b0\vert^2)^{3/2}} + \frac{2\vert\xi_\b0\vert(h_z+\omega_\b0)}{\vert h_+\vert} &=& 1-\sideset{}{^\prime}\sum_\bq n_\bq/N.~~~~~\label{eq:xi0}
}
The derivation of Eqs.~(\ref{eq:damping})--(\ref{eq:xi0}) are given in Appendix~\ref{sec:app}. Since the precession frequency shift due to anisotropies and demagnetization are already captured by the effective field $h_z$,\cite{kittel1948} we set $\omega_\b0$ to zero and obtain
\eq{
  \dot{\theta} e^{i\phi} &=& ih_+\big(1-n_{tot}/N\big) + \xi_\b0\Gamma,\label{eq:precession-rate}
  \\\xi_\b0/h_+ &=& \left\{
  \begin{array}{ll}
    0.5/h_z, & ~\vert h_+\vert \ll +h_z.\\
    0.6/\vert h_+\vert,  & ~\vert h_+\vert \gg \vert h_z\vert.\\
    1.0/\vert h_+\vert, & ~\vert h_+\vert \ll -h_z.\\
  \end{array}\right.\label{eq:gamma-monotonic}
}
Equation~(\ref{eq:precession-rate}) shows that the frequency of precession decreases with the total magnon population $n_{tot}$, and is therefore strongly dependent on temperature.\cite{Qi2007} We have neglected the frequency detuning due to the $W_\bq$ scattering term [see text after Eq.~(\ref{eq:A4})] and instead focus on the damping effect that causes the magnetization to spiral toward the field. In a fixed reference frame, Eq.~(\ref{eq:precession-rate}) reproduces the Landau-Lifshitz (LL) equation $\partial_t\vec{M}=\vec{M}\times(\vec{h}-\alpha\hat{M}\times\vec{h})$ for $n_{tot}=0$, with $\alpha=\vert\xi_\b0\Gamma/h_+\vert$. Thus, the phenomenological damping parameter $\alpha$ in the LL equation is in fact not constant, but dependent on the angle $\theta_h$ between the magnetization and the field $\vec{h}$. The LL equation grossly underestimates the true damping when $\theta_h$ is close to $\pi$. Since $\alpha$ also depends on $W_\bq$, a microscopic derivation of $W_\bq$ would be necessary for studying the relaxation physics beyond the LL theory. We will not go into such details here,\cite{Suhl2007} but instead examine the magnetization dynamics resulting from specific field-dependence of $\alpha$ in the next section.

The delta function in Eq.~(\ref{eq:energy-conservation}) indicates energy conservation when exciting a magnon mode. The nonlinear frequency renormalization in Eq.~(\ref{eq:freq-shift}) does not appear in the semiclassical treatment of the ferromagnetic resonances. For a Heisenberg ferromagnet in a magnetic field, the uniform mode lies at the minimum of the magnon dispersion curve and therefore, has no relaxation. Anderson and Suhl showed that magnetic dipole interaction causes the magnon dispersion curve to split into a band which depends on the magnitude and direction of the magnon wave vector,\cite{Anderson1955} thus creating some phase space for satisfying the constraint $\omega_\b0=\omega_\bq$.\cite{Clogston1956} This degenerate magnon mechanism, as it was known, does not work for a magnetic thin film when the magnetization has a significant out-of-plane component.\cite{Arias1999} Equations~(\ref{eq:freq-shift}) and (\ref{eq:xi0}) show that an additional shift of $O(\vert\vec{h}\vert\sin^2\theta_h)$ or $O(\vert\vec{h}\vert)$ is present respectively for $\theta_h$ less than or greater than $\pi/2$, where $\theta_h$ is the angle between $\vec{h}$ and the Berger axis. This was not found earlier since the field direction is chosen as the $z$-axis in conventional HP expansion ($\theta_h=0$) to eliminate the linear term in $b_\b0$.

%%%%%%%%%%%%%%%%%%%%%%%%%%%%%%%%%%%%%%%%%%%%%%%%%%%%%%%%%%%%%%%%%%%%%%%%%%%%%%%%%
%%%%%%%%%%%%%%%%%%%%%%%%%%%%%%%%%%%%%%%%%%%%%%%%%%%%%%%%%%%%%%%%%%%%%%%%%%%%%%%%%
\subsection{STT-driven dynamics in a field}\label{sec:numerics}

With a quantum approach for modeling STT-driven magnetization dynamics in an effective magnetic field, we describe the implementation details for a Monte Carlo simulation for the study of STT-driven precession and magnetization switching in a magnetic film.

%%%%%%%%%%%%%%%%%%%%%%%%%%%%%%%%%%%%%%%%%%%%%%%%%%%%%%%%%%%%%%%%%%%%%%%%%%%%%%%%%
\subsubsection{Description of Monte Carlo simulation}

Consider the current as a stream of electrons incident on the magnetic film from one side at a uniform rate and constant momentum, which is a reasonable simplification due to insensitivity of the noise to the electron momentum distribution found in Ref.~\onlinecite{Wang2012}. For each incident electron, we compute the scattering amplitudes for each of four possible elastic outcomes as well as the numerous inelastic outcomes. Occupation numbers for the electron levels in the normal metals are sampled stochastically according to their Fermi distributions and forbidden outcomes are then projected out as required by the Pauli exclusion principle. Next, we sample a scattering outcome and determine the new orientation of the magnetization in the case of an elastic scattering, or update the internal energy of the magnet for an inelastic outcome. The magnetization orientation is then allowed to precess under the effective field until the next electron impinges on the film. We assume that the magnet is in quasi-equilibrium in between successive electron scatterings, and the temperature of the magnet is updated by requiring that the new total energy of the magnet is equal to that given by the Bose-Einstein distribution of magnons in the magnet.

In the following, we illustrate our approach by applying it to a magnet in various fields. We consider a magnet comprising a $100\times 100\times 10$ spin-1/2 moments with ${\cal J}=80$~meV and $\lambda_0=-0.7$~eV. All incident electrons have the same momentum $k_x=9~{\rm nm}^{-1}$ and an average spin polarization of $25\%$ with respect to a certain direction. The coupling strength between an itinerant electron and the spin-1/2 moments in the magnet is taken to be $\lambda=1.2$~eV. In principle, we should implement Landauer's principle numerically by sampling electrons from both sides of the film according to their respective Fermi distributions. But this is computationally very costly and we leave that for future studies where the noise in a much smaller magnet is instead the main focus.

%%%%%%%%%%%%%%%%%%%%%%%%%%%%%%%%%%%%%%%%%%%%%%%%%%%%%%%%%%%%%%%%%%%%%%%%%%%%%%%%%
\subsubsection{Effective field}

In the Berger frame, the effective field experienced by a single-domain magnet, due to the Zeeman energy $-\vec{h}_B\cdot\bJ$, an easy-axis anisotropy energy $h_K\sin^2\theta_K$, and an easy-plane anisotropy energy $h_P\cos^2\theta_P$, is
\eq{
  \vec{h}=\vec{h}_B + \vec{h}_K\cos\theta_K - \vec{h}_P\cos\theta_P,
}
where $\vec{h}_K$ and $\vec{h}_P$ respectively lie along the easy-axis and the normal to the easy-plane, $\theta_K$ is the angle between $\vec{h}_K$ and $\bJ$, and $\theta_P$ is the angle between $\vec{h}_P$ and $\bJ$. The demagnetization field of a thin magnetic film can be lumped into $h_P$. For dimensionless spin operator $\bJ$, each field is given in units of energy. Thus, 1~kOe would correspond to $11.6$~$\mu$eV for the electron gyromagnetic ratio $g=2$.

Let us define the Berger frame such that its $z$-axis is specified by $(\theta_M,\phi_M)$ with respect to a fixed reference frame, and its $y$-axis lies along $\hat{y}\cos\phi_M-\hat{x}\sin\phi_M$ in the fixed frame. Using Eq.~(\ref{eq:precession-rate}) to obtain the new magnetization direction $(\theta,\phi)$ in the initial Berger frame, the new orientation $(\theta'_M,\phi'_M)$ in the fixed frame is given by
\eq{
  \cos\theta'_M &=& \cos\theta_M\cos\theta - \cos\phi\sin\theta_M\sin\theta,
  \\\phi'_M &=& \phi_M + \arg\Big[\cos\theta\sin\theta_M \nonumber
    \\&& + \sin\theta\left(\cos\theta_M\cos\phi + i\sin\phi\right)\Big].
}
Note that the Berry phase associated with the rotating Berger frame is implicitly captured by these equations.

%%%%%%%%%%%%%%%%%%%%%%%%%%%%%%%%%%%%%%%%%%%%%%%%%%%%%%%%%%%%%%%%%%%%%%%%%%%%%%%%%
\subsubsection{Magnetic field only}

\begin{figure}
  \centering
  \includegraphics[width=\columnwidth]{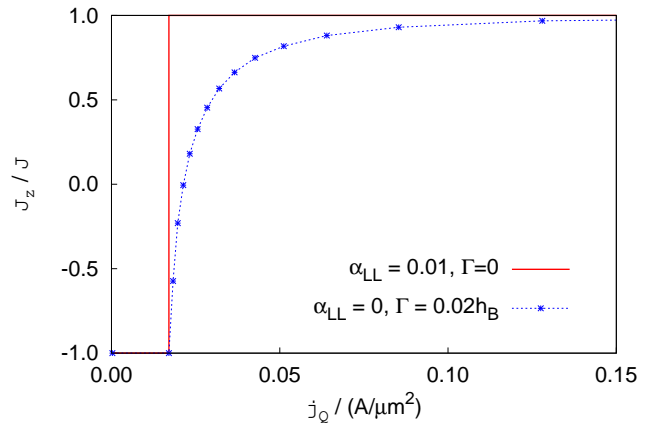}
  \caption{$J_z/J$ versus current for an STT-driven precession in a vertical magnetic field with damping $\alpha=\alpha_{LL}+\vert\xi_\b0\Gamma/h_+\vert$. The LL damping ($\alpha_{LL}=0.01$, $\Gamma=0$) results in an abrupt transition between no-switching and complete-switching at a critical current density $0.017~{\rm A}$/$\mu{\rm m}^2$, while the damping with $\alpha_{LL}=0$ and $\Gamma=0.02h_B$ shows a gradual switching for current density above the critical value.}
  \label{fig:zeeman}
\end{figure}

Consider an external field with $h_B=1$~kOe along the negative $z$-axis and a current with average spin polarization along the $z$-direction. With this neglect of the demagnetization field and all anisotropies, Fig.~\ref{fig:zeeman} compares the $J_z$ component of the precessing magnetization for different damping $\alpha$, with $\alpha=\alpha_{LL}+\vert\xi_\b0\Gamma/h_+\vert$. For Landau-Lifshitz damping with $\alpha_{LL}=0.01$ and $\Gamma=0$, complete switching occurs for any current above $0.017~{\rm A}$/$\mu{\rm m}^2$ (Landau-Lifshitz-Gilbert equation would give similar result since $\alpha$ is small). For $\alpha_{LL}=0$ and $\Gamma=0.02h_B$ (the magnitude of $\Gamma$ is chosen to give the same critical current), $J_z$ switches progressively after the current crosses the critical value.

%%%%%%%%%%%%%%%%%%%%%%%%%%%%%%%%%%%%%%%%%%%%%%%%%%%%%%%%%%%%%%%%%%%%%%%%%%%%%%%%%
\subsubsection{Easy-plane anisotropy only}

\begin{figure}
  \centering
  \includegraphics[width=\columnwidth]{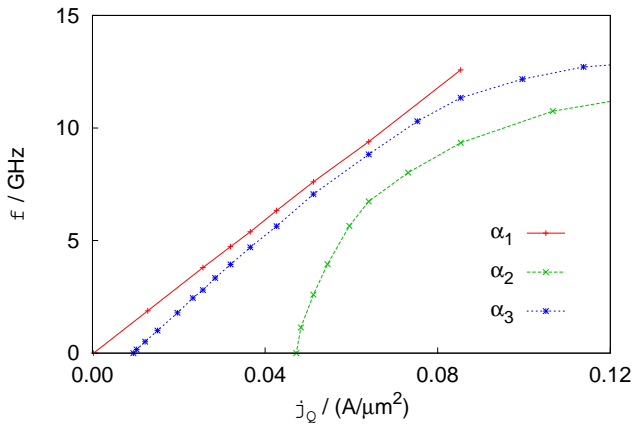}
  \caption{Precession frequency against current through a magnetic film (with electron spin polarization normal to the film), for damping $\alpha=\alpha_{LL}+\vert\xi_\b0\Gamma/h_+\vert$. The film contains easy-plane anisotropy but no other anisotropy or field is present. The precession frequency for $\alpha_1$ damping ($\alpha_{LL}=0.01$, $\Gamma=0$) is proportional to the current, while the $\alpha_2$ damping ($\alpha_{LL}=0$, $\Gamma=0.01h_P$) imposes a threshold current for exciting a precession, and the $\alpha_3$ damping ($\alpha_{LL}=0.008$, $\Gamma=0.002h_P$) lies between the two. The precession frequency saturates at $13.1$~GHz.}
  \label{fig:planar}
\end{figure}

Here, we consider a magnetic film with an easy-plane anisotropy $h_P=5$~kOe (inclusive of a contribution from the demagnetization field), but no other anisotropy or field is present. When a current with spin polarization along the hard-axis passes through the film, an STT-driven precession occurs with a frequency that increases with the out-of-plane magnetization. Figure~\ref{fig:planar} compares the precession frequency for different damping $\alpha$, with $\alpha=\alpha_{LL}+\vert\xi_\b0\Gamma/h_+\vert$. For the LL damping ($\alpha_{LL}=0.01$, $\Gamma=0$) the precession frequency is proportional to the current, indicating that the out-of-plane magnetization is proportional to the current. This would be observed if the Gilbert damping from adiabatic spin pumping effect of Ref.~\onlinecite{Tserkovnyak2002a} is much greater than impurity-originated scattering between the uniform mode and the magnons. The damping with $\alpha_{LL}=0$ and $\Gamma=0.01h_P$ imposes a threshold current for exciting a uniform precession and a large current is required for complete switching. In experiments, one might perhaps observe the third case where $\alpha_{LL}=0.008$ and $\Gamma=0.002h_P$ are chosen for illustration.

%%%%%%%%%%%%%%%%%%%%%%%%%%%%%%%%%%%%%%%%%%%%%%%%%%%%%%%%%%%%%%%%%%%%%%%%%%%%%%%%%
\subsubsection{Thin film with in-plane uniaxial and magnetic fields}

\begin{figure}
  \centering
  \begin{tabular}{c}
    \subfigure[]{
      \includegraphics[trim=70 65 70 50, scale=1]{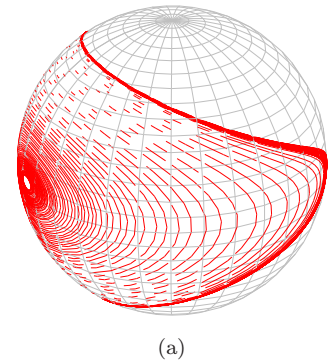}
      \label{fig:precession}
    }\\
    \subfigure[]{
      \includegraphics[trim=0 0 0 5, width=\columnwidth]{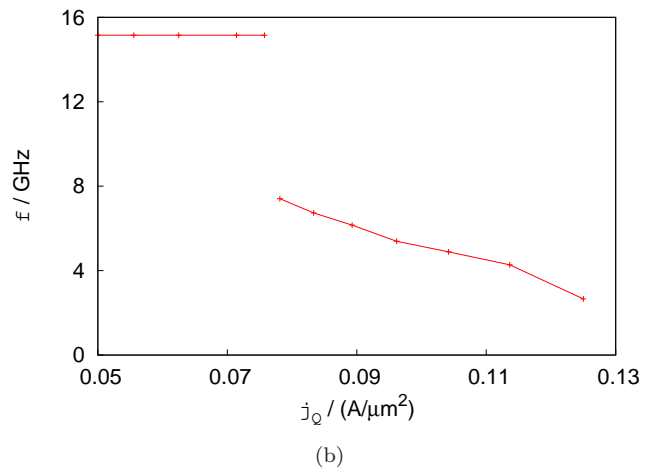}
      \label{fig:switching-Bfield}
    }
  \end{tabular}
  \caption{(a) Magnetization trajectory driven by a current density $j_Q=0.1~{\rm A}$/$\mu{\rm m}^2$. The orientation spirals away from its initial direction before reaching a steady orbit. (b) For $j_Q<0.078~{\rm A}$/$\mu{\rm m}^2$, the magnetization undergoes a small-amplitude precession at a constant frequency 15.2~GHz. Above this critical current, the trajectory jumps to a large, distorted orbit with a precession frequency that decreases with the current.}
\end{figure}

Lastly, we consider a magnetic film with $h_P=7.0$~kOe along the $z$-axis, $h_K=1.2$~kOe along the $x$-axis, and an in-plane magnetic field $h_B=2.0$~kOe applied at $5^\circ$ to the easy-axis. Using $\Gamma=0.8~\mu$eV, Fig.~\ref{fig:precession} shows a Monte Carlo simulation of the magnetization trajectory driven by a current density $j_Q=0.1~{\rm A}$/$\mu{\rm m}^2$. The strong easy-plane anisotropy results in a highly distorted trajectory. Figure~\ref{fig:switching-Bfield} shows the dependence of the precession frequency on the current. Below the critical current density $0.078~{\rm A}$/$\mu{\rm m}^2$, the magnetization undergoes a small-amplitude precession with a constant frequency 15.2~GHz. Above this critical current, the trajectory jumps to a large, distorted orbit with a precession frequency that decreases rapidly as the current increases further. This qualitatively reproduces the classical Landau-Lifshitz-Gilbert macrospin simulation in the experimental study of Ref.~\onlinecite{Kiselev2003}.

%%%%%%%%%%%%%%%%%%%%%%%%%%%%%%%%%%%%%%%%%%%%%%%%%%%%%%%%%%%%%%%%%%%%%%%%%%%%%%%%%
%%%%%%%%%%%%%%%%%%%%%%%%%%%%%%%%%%%%%%%%%%%%%%%%%%%%%%%%%%%%%%%%%%%%%%%%%%%%%%%%%
%%%%%%%%%%%%%%%%%%%%%%%%%%%%%%%%%%%%%%%%%%%%%%%%%%%%%%%%%%%%%%%%%%%%%%%%%%%%%%%%%
%%%%%%%%%%%%%%%%%%%%%%%%%%%%%%%%%%%%%%%%%%%%%%%%%%%%%%%%%%%%%%%%%%%%%%%%%%%%%%%%%
\section{Discussions and conclusion}\label{sec:conclusion}

We have presented a simple microscopic model in a given field, together with contact exchange interaction between an itinerant electron and the spins in the magnet and detailed the quantum scattering between the current electron and the rigid macrospin states and the excited states (magnons) of the nanomagnet which produces SST-driven precession  and switching. The quantum approach in the large $J$ limit produces the current electron spin rotation which is used with angular momentum conservation to infer the macrospin recoil of Slonczewski\cite{Slonczewski1996} and reproduces the zero-wave vector spin wave emission of Berger.\cite{Berger1996} By contrast, the quantum method deduces the recoil dynamics directly without the explicit use of angular momentum conservation. The use of angular momentum conservation for the spin torque transfer is valid where there is a large number of incoherent degrees of freedom involved but may be broken in approaching the quantum limit through uncertainty created by either ultrafast control or atomic scale spin localization.  Our simple simulations demonstrate the capability of the quantum approach to study damped magnetization dynamics starting from a microscopic treatment of ferromagnetic relaxation, the noise sources from the transverse spin fluctuations in unitary scattering and energy transfer from the inelastic scattering involving magnons.

We note that an important difference exists between our Holstein-Primakoff boson expansion and existing studies in the literature. The latter often expand the spin operators about a fixed magnetic field, thus restricting studies of magnon relaxation to small angles between the magnetization direction and the applied magnetic field. In constrast, our expansion in Eqs.~(\ref{eq:magnon-magnon}) and (\ref{eq:Hamiltonian-in-field}) does not require the magnetization direction to be close to a fixed axis. We showed that the electron-macrospin interaction in our HP expansion exactly reproduces Wang and Sham's quantum correction at the leading $1/\sqrt{N}$ order. Furthermore, the HP model is able to treat precession damping even at large angles from the field.

We have treated the current electrons  as uncorrelated with one another in the scattering events with the magnet  which provide a stochastic source for the STT-driven magnetization dynamics. In our simulations, we left out a few aspects that are to be treated quantum-mechanically. First, fluctuations in the macrospin direction in connection to the fluctuation-dissipation theorem for macrospin-magnon scatterings in a field had not been included. Second, the macrospin recoil due to adiabatic spin pumping effect of Ref.~\onlinecite{Tserkovnyak2002b} was also excluded. Third, contributions from electron-magnon scatterings to relaxation in ferromagnetic metals had not been considered. Together, these can be taken into consideration by replacing the damping term with a phenomenological Wiener process with drift, or one might try to extend our quantum approach to include them. All these would be necessary for simulating the ``thermally-activated switching''.\cite{Koch2000} Our solution of the magnetization dynamics also has to be extended to apply to more coherent sources of the driving mechanism, such as microwave,\cite{Kiselev2003} laser,\cite{Kirilyuk2010}  and external spin waves.\cite{Wang2011} The quality of the quantum approach also has to be tested for nano-magnets or magnetic dots in the areas of spin pumping in metallic\cite{Kiselev2003} or insulating ferromagnets,\cite{Tserkovnyak2002b} magnonics\cite{Kruglyak2010b}, and domain wall dynamics,\cite{Hayashi2012} in which the magnetization had hithertofore  be treated as classical using micromagnetics.

%%%%%%%%%%%%%%%%%%%%%%%%%%%%%%%%%%%%%%%%%%%%%%%%%%%%%%%%%%%%%%%%%%%%%%%%%%%%%%%%%%%%%%
%%%%%%%%%%%%%%%%%%%%%%%%%%%%%%%%%%%%%%%%%%%%%%%%%%%%%%%%%%%%%%%%%%%%%%%%%%%%%%%%%%%%%%
%%%%%%%%%%%%%%%%%%%%%%%%%%%%%%%%%%%%%%%%%%%%%%%%%%%%%%%%%%%%%%%%%%%%%%%%%%%%%%%%%%%%%%
%%%%%%%%%%%%%%%%%%%%%%%%%%%%%%%%%%%%%%%%%%%%%%%%%%%%%%%%%%%%%%%%%%%%%%%%%%%%%%%%%%%%%%
\acknowledgments

The research is supported by the NSF through grant DMR-1124601 and ECCS-1202583. Tay thanks Xiaofeng Shi for useful discussions.

%%%%%%%%%%%%%%%%%%%%%%%%%%%%%%%%%%%%%%%%%%%%%%%%%%%%%%%%%%%%%%%%%%%%%%%%%
%%%%%%%%%%%%%%%%%%%%%%%%%%%%%%%%%%%%%%%%%%%%%%%%%%%%%%%%%%%%%%%%%%%%%%%%%
%%%%%%%%%%%%%%%%%%%%%%%%%%%%%%%%%%%%%%%%%%%%%%%%%%%%%%%%%%%%%%%%%%%%%%%%%
%%%%%%%%%%%%%%%%%%%%%%%%%%%%%%%%%%%%%%%%%%%%%%%%%%%%%%%%%%%%%%%%%%%%%%%%%
\appendix
\section{Derivation of Eqs.~(\ref{eq:damping})--(\ref{eq:xi0})}\label{sec:app}

First, we substitute $\eta_\b0 = \eta'_\b0+\bar{\eta}_\b0$ into Eq.~(\ref{eq:differentialEq}), with constant $\bar{\eta}_\b0$, and obtain
\eq{
  i\partial_t \eta'_\b0 &=& \Big(h_z+\omega_\b0 + \frac{h_+\bar{\eta}^*_\b0}{2\sqrt{N}} + \frac{h_-\bar{\eta}_\b0}{2\sqrt{N}}\Big)\eta'_\b0 \nonumber
  \\ &-& \frac{\sqrt{N}h_+}{2}\Big(1-\sideset{}{^\prime}\sum_\bq n_\bq/N\Big) + \frac{h_-}{4\sqrt{N}}\sideset{}{^\prime}\sum_\bq \eta_\bq \eta_{-\bq} \nonumber
  \\&+& (h_z+\omega_\b0)\bar{\eta}_\b0 + \frac{h_+}{2\sqrt{N}}(\eta'^*_\b0\eta'_\b0 + \bar{\eta}^*_\b0\bar{\eta}_\b0 + \bar{\eta}_\b0\eta'^*_\b0) \nonumber
  \\&+& \frac{h_-}{4\sqrt{N}}(\eta'^2_\b0 + \bar{\eta}^2_\b0) + \frac{1}{\sqrt{N}}\sideset{}{^\prime}\sum_\bq W^*_\bq \eta_\bq.\label{eq:A1}
}
Note the grouping of terms linear in $\eta'_\b0$ that contribute to a renormalized energy for the uniform mode. The remaining terms are either approximately constant or have phases that rotate with frequencies far from that of $\eta'_\b0$. Since $\eta_\b0=0$ in the Berger frame at $t=0$, we further substitute $\eta'_\b0=-\bar{\eta}_\b0=-\sqrt{N}h_+\tau_\b0$ (with real $\tau_\b0$) into the last two lines of Eq.~(\ref{eq:A1}) and obtain
\eq{
  i\partial_t \eta'_\b0 &=& \sqrt{N}h_+\Big\{\vert h_+\vert^2\tau^2_\b0 + (h_z+\omega_\b0)\tau_\b0 - \frac{1}{2} + \sideset{}{^\prime}\sum_\bq \frac{n_\bq}{2N}\Big\} \nonumber\label{eq:A2}
  \\&+& \left(h_z+\omega_\b0 + \vert h_+\vert^2\tau_\b0\right)\eta'_\b0  + \frac{1}{\sqrt{N}}\sideset{}{^\prime}\sum_\bq W^*_\bq \eta_\bq.
}
We have neglected $\eta_\bq\eta_{-\bq}$ since this gives a higher order correction to damping compared to the $W_\bq$ term. We now eliminate the strong driving term on the first line of Eq.~(\ref{eq:A2}) by an appropriate choice of $\tau_\b0$. Defining $\eta'_\b0 = \tilde{\eta}_\b0 e^{-i(h_z+\tilde{\omega}_\b0)t}$ and $\eta_\bq=\tilde{\eta}_\bq e^{-i\omega_\bq t}$, where $\tilde{\omega}_\b0=\omega_\b0+\vert h_+\vert^2\tau_\b0$, we use the Weisskopf-Wigner approximation to obtain\cite{Scully1997}
\eq{
  \partial_t\tilde{\eta}_\b0 &=& -\frac{\tilde{\eta}_\b0}{N}\sideset{}{^\prime}\sum_\bq\vert W_\bq\vert^2 \int^t_0 dt' e^{i(\tilde{\omega}_\b0-\omega_\bq)t'}, \label{eq:Weisskopf}
  \\&\approx& -\tilde{\eta}_\b0\Gamma/2,\label{eq:A4}
}
where $\Gamma$ is defined in Eq.~(\ref{eq:energy-conservation}). The approximation made in Eq.~(\ref{eq:A4}) consists of extending the upper limit of the integral to infinity and neglecting the principal part of $\int^\infty_0 dt~e^{i\omega t}=iP(1/\omega)+\pi\delta(\omega)$. Transforming back to $\eta_\b0$ then gives Eq.~(\ref{eq:damping}), where $\xi_\b0$ in the main text is related to $\tau_\b0$ via $\xi_\b0=h_+\tau_\b0$.

When the angle $\theta_h$ between the magnetization and the field is close to $\pi$, Eq.~(\ref{eq:A2}) gives a divergence in $\xi_\b0$, thus indicating that the truncated HP expansion at cubic order is no longer valid. To remedy this, we include every term in the infinite HP series that contains only $\eta_\b0$ or $\eta^*_\b0$. This gives
\eq{
  i\partial_t \eta_\b0 &=& \frac{1}{\sqrt{N}} \sum_{\bq\neq\b0} W^*_\bq \eta_\bq \nonumber
  \\&+& \frac{\sqrt{N}h_+}{2}\left\{-1+\sum^\infty_{n=1}\frac{n+1}{2^n}\frac{(2n-3)!!}{n!}\frac{\eta^{*n}_\b0\eta^n_\b0}{N^n}\right\} \nonumber
  \\&+& \frac{\sqrt{N}h_-}{2}\sum^\infty_{n=1}\frac{n}{2^n}\frac{(2n-3)!!}{n!}\frac{\eta^{*n-1}_\b0\eta^{n+1}_\b0}{N^n} \nonumber
  \\&+& \frac{h_+}{2\sqrt{N}}\sideset{}{^\prime}\sum_\bq\eta^*_\bq\eta_\bq +  (h_z+\omega_\b0) \eta_\b0,
}
where $n!!=n(n-2)\cdots l$, $l=1,2$ for odd and even $n$ respectively, and $n!!=1$ for $n\leq 0$. As before, we substitute $\eta_\b0 = \eta'_\b0+\bar{\eta}_\b0$ and $\bar{\eta}_\b0=\sqrt{N}\xi_\b0=\sqrt{N}h_+\tau_\b0$, and obtain
\eq{
  i\partial_t \eta'_\b0 &=& \frac{\sqrt{N}h_+}{2}\Big\{-1 + \sum^\infty_{n=1}\frac{n(n+1)}{2^{n-1}}\frac{(2n-3)!!}{n!}\vert\xi_\b0\vert^{2n}\Big\} \nonumber
}
\eq{
  &+& \frac{h_+}{2\sqrt{N}}\sideset{}{^\prime}\sum_\bq\eta^*_\bq\eta_\bq + \left(h_z+\tilde{\omega}_\b0\right)\eta'_\b0 \nonumber
  \\&+& \frac{1}{\sqrt{N}}\sideset{}{^\prime}\sum_\bq W^*_\bq \eta_\bq,\\
  \tilde{\omega}_\b0 &=& \omega_\b0 + \sum^\infty_{n=1}\frac{n(n+1)}{2^n}\frac{(2n-3)!!}{n!}\frac{\vert\xi_\b0\vert^{2n}}{\tau_\b0}.
}
By differentiating $\sqrt{1-x}$ and its series expansion, we obtain the following identity,
\eq{
  \sum^\infty_{n=1} \frac{n(n+1)}{2^n}\frac{(2n-3)!!}{n!}x^n \equiv \frac{x(4-3x)}{(1-x)^{3/2}},
}
which then gives
\eq{
  i\partial_t\eta'_\b0 &=& (h_z+\tilde{\omega}_\b0)\eta'_\b0 + \frac{1}{\sqrt{N}}\sideset{}{^\prime}\sum_\bq W^*_\bq \eta_\bq,
}
with $\tilde{\omega}_\b0$ and $\xi_\b0$ satisfying Eqs.~(\ref{eq:freq-shift}) and (\ref{eq:xi0}) respectively. One can show that $\tilde{\omega}_\b0\geq\omega_\b0$ holds strictly.

\bibliography{STT.bbl}

\end{document}